\documentclass[aps,preprint,superscriptaddress,showpacs,12pt]{revtex4-1}
\usepackage{amsmath}
\usepackage{dcolumn}
\usepackage{comment}
\usepackage[dvips]{graphicx}
\begin{document}
\title{Precise measurement of coupling strength and high temperature quantum effect in
a nonlinearly coupled qubit-oscillator system}
\author{Li Ge}
\affiliation{School of Science, Hangzhou Dianzi University, Hangzhou, 310018, China}
\affiliation{Beijing Computational Science Research Center, Beijing, 100084, China}
\author{Nan Zhao}
\email{nzhao@csrc.ac.cn}
\affiliation{Beijing Computational Science Research Center, Beijing, 100084, China}

\begin{abstract}

We study the coherence dynamics of a qubit coupled to a harmonic oscillator with both linear and quadratic interactions.
As long as the linear coupling strength is much smaller than the oscillator frequency,
the long time behavior of the coherence is dominated by the quadratic coupling strength $g_2$.
The coherence decays and revives at a period $2\pi/g_2$, with the width of coherence peak decreases as the temperature increases,
hence providing a way to measure $g_2$ precisely without cooling. Unlike the case of linear coupling, here the coherence dynamics
never reduces to the classical limit in which the oscillator is classical. Finally, the validity of linear coupling approximation is discussed
and the coherence under Hahn-echo is evaluated.


\end{abstract}

\pacs{03.65.Yz, 05.40.Ca, 42.50.Lc}

\maketitle

Coherence is one of the most remarkable features that distinguishes a quantum system with its classical counterpart.
While it does not change for an isolated system, the coherence of an open system usually decays due to the interaction with a heat bath~\cite{zurek1}.
Many efforts have been made to understand this decoherence process~\cite{everett, lindblad, leggett1, leggett2, unruh, myatt,zurek2,weiss},
with the purpose to suppress it~\cite{zanardi,shor,ekert, lloyd1, lloyd2} or to detect signals from the bath~\cite{zhao1,lukin1,hanson,lavrik,ganzhorn,chaste,jensen,naik,zhao2}.
Among the numerous theoretical models, the pure dephasing model is special since it admits no dissipation but only dephasing.
Typically, a pure dephasing model describes a qubit linearly coupled to the bath by the interaction: $H_I=g_1\sigma_zx$,
where $\sigma_z$ is the qubit operator and $x$ the bath variable. This model has been extensively studied
~\cite{uhrig,lidar,sarma,uys,liu,viola,itano}, in the case that $x$ obeys Gaussian statistics
the coherence is given by a simple expression in terms of the correlation function of $x$.


While many efforts are devoted to the linear coupling model, it is just an approximation in most cases. Generally, the pure-dephasing type
of coupling takes the form $H_I=\sigma_z f(x)$, where $f(x)$ is a function of the environment variable, as realized
in the systems of superconducting qubit or semiconductor quantum dot~\cite{tsai,shnirman1,shnirman2,vion,medford}. Expanding $f(x)$ to first,
second...orders gives linear, quadratic...couplings, usually the higher order coupling strengths are much smaller than $g_1$
so that one can make the linear approximation. However for a superconducting qubit, the value of $g_1$
can be easily tuned and even be $0$ at an optimal point~\cite{tsai,shnirman1},
then the quadratic coupling is necessarily dominant. This raises theoretical interest in quadratic coupling,
and efforts have been made to understand the decoherence of qubit in such case by using the Bloch-Redfield approach
 or the linked-cluster expansion (LCE)~\cite{ithier,shnirman3,lukasz}.
In these approaches, the bath is totally characterized by the noise power spectrum,
while its own dynamics is not specified. Unlike that of linear coupling,  the LCE does not stop
at the lowest order, and the higher order terms represent non-classical noise.

In this paper we consider a qubit-oscillator system with both linear and quadratic coupling,
and $x$, as the coordinate of the oscillator, is a dynamical variable. Such model can be realized, e.g., in some devices
of quantum nondemolition measurements consisting of a flux qubit and a LC-resonator~\cite{luca,wei,reuther}, or in a
hybrid system of a magnetized mechanical resonator and a single electron spin associated with a nitrogen
vacancy center in diamond~\cite{lukin2}. In the latter case, the mechanical motion of the resonator is coupled to the electron spin by an
interaction $H_I=(E_0+g_1x+\frac{1}{2}g_2x^2)\sigma_z$, where $E_0$ is the Zeeman energy  produced by
the magnetic field of the resonator in its equilibrium position, $x$ the displacement of the resonator and $g_1$, $g_2$
proportional to the first and second order derivatives of the magnetic field, respectively.
Recently, Zhao and Yin~\cite{zhao2} pointed that by measuring the qubit coherence in a linear coupling qubit-oscillator system,
the oscillator frequency or the linear coupling strength can be determined with high precision at room temperature. So a natural question is
whether the (usually weaker) quadratic coupling can result in some remarkable effect and in turn, be measured precisely? It is also known
that in the linear model the oscillator behaves classically if the temperature is much larger than its frequency,
so will the quadratic coupling enhance the quantum effect?

In the following we show that in the case of free evolution, the qubit coherence decays and revives at a period $2\pi/g_2$.
The width of the coherence peak decreases as the temperature increases, so its position can be measured precisely in high temperature.
The coherence dynamics is quite different from that of linear coupling, in the sense that it never
reduces to the classical limit, and the reason for this is figured out. Since the quadratic coupling is present in many systems,
our result clarifies the conditions under which the linear approximation is valid.
The evolution of coherence under Hahn echo is also presented in the end.

\section{Calculation of the qubit coherence}
We first consider the qubit coupled to a single model oscillator, which is described by the Hamiltonian:
\begin{equation}
H_s=H_q+H_0+H_I=\frac{1}{2}\omega_q\sigma_z+\frac{1}{2}\omega_0(\hat{x}^2+\hat{p}^2)+g_1\sigma_z \hat{x}+\frac{1}{2}g_2\sigma_z \hat{x}^2.
\end{equation}
Throughout this paper $\hat{x}$ and $\hat{p}$ are quantum operators and $x$, $p$ are classical variables.
We set $\hbar=1$ and rescale $\hat{x}\rightarrow \frac{\hat{x}}{m\omega_0}$, $\hat{p}\rightarrow \hat{p}m\omega_0$ so that the mass
does not appear in the Hamiltonian. In the following we focus on the case $g_1\ll\omega_0$ and $g_2\ll\omega_0$, this is the parameter
regimes achieved in experiments~\cite{lukin2}, where $g_1/\omega_0< 10^{-2}$ , and $g_2$ is neglected.

The qubit is initially prepared in a superposition state $|\chi\rangle = \left(|0\rangle + | 1\rangle\right)/\sqrt{2}$,
where  $|0\rangle$ and $|1\rangle$ are eigenbases of the qubit corresponding to $\sigma_z=-1$ and $+1$, respectively.
Since $\sigma_z$ is a conserved, we focus on the dynamics of the relative phase, or
the quantum coherence, between qubit states $\vert 0\rangle$ and $\vert 1\rangle$.
If the oscillator is initially in a pure state $|\varphi\rangle$, the whole system will evolve as
$e^{-iH_st} |\chi\rangle \otimes|\varphi\rangle=\frac{1}{\sqrt{2}}e^{i\frac{\omega_q}{2}}|0\rangle \otimes e^{-iH_-t}|\varphi\rangle+
\frac{1}{\sqrt{2}}e^{-i\frac{\omega_q}{2}}|1\rangle \otimes e^{-iH_+t}|\varphi\rangle$, where $H_{\pm}=H_0\pm(g_1 \hat{x}+\frac{1}{2}g_2\hat{x}^2)$.
Then the coherence decays with a factor $L=\langle\varphi| e^{iH_{-}t}e^{-iH_{+}t} |\varphi\rangle$
(the unimportant phase $e^{-i\omega_qt}$ is neglected). This formula can be generalized to the case where the oscillator is initially in
a mixed state described by a density matrix $\rho_0$~\cite{yang}:
\begin{equation}
L = Tr(e^{-iH_{+}t}\rho_0e^{iH_{-}t})
\end{equation}
Commonly $\rho_0$ is the thermal state: $\rho_0=\frac{1}{Z}e^{-\beta H_0}$, with the partition function $Z=1/(2\sinh\frac{\beta \omega_0}{2})$,
and $\beta=1/k_BT$ is the inverse temperature.
For $g_2=0$, the tracing is easily done in the Fock space~\cite{breuer}: $a^{\dagger}a|n\rangle=n|n\rangle$,
with $a=\sqrt{\frac{1}{2}}(\hat{x}+i\hat{p})$ the annihilation operator.
However, this method turns out to be inefficient in the presence of quadratic coupling, also the commonly used LCE
encounters great difficulties. We note here $H_0$, $H_{+}$, $H_{-}$ are all quadratic forms of $\hat{x}$ and $\hat{p}$, so their
propagators can be evaluated exactly in the coordinate basis $|x\rangle$, in which the coherence is:
\begin{equation}
L =\frac{1}{Z}\int dx_1 dx_2 dx_3 \langle x_1|e^{-\beta H_0}|x_2\rangle\langle x_2|e^{iH_{-}t}|x_3\rangle \langle x_3|e^{-iH_{+}t}|x_1\rangle
\end{equation}
The propagators are:
\begin{equation}
\langle x_3|e^{-iH_{+}t}|x_1\rangle
=\sqrt{\frac{\omega_1}{2 \pi i\omega_0 \sin \omega_1t}}\exp\{\frac{i\omega_1}{2\omega_0\sin\omega_1t}[(x_1^2+x_3^2)\cos\omega_1t-2x_1x_3]+J_1(x_1+x_3)\}, \nonumber \\
\end{equation}
where $\omega_1=\sqrt{\omega_0(\omega_0+g_2)}$, $J_1=\frac{ig_1}{\omega_1}(\cot\omega_1t-\frac{1}{\sin\omega_1t})$.
$\langle x_3|e^{-iH_{-}t}|x_2\rangle$ and $\langle x_1|e^{-\beta H_0}|x_2\rangle$ have similar forms with $\omega_1$ and $J_1$
replaced by $\omega_2=\sqrt{\omega_0(\omega_0-g_2)}$ and $J_2=\frac{ig_1}{\omega_2}(\cot\omega_2t-\frac{1}{\sin\omega_2t})$.

The propagators $\langle x_3|e^{-iH_{+}t}|x_1\rangle$ and $\langle x_3|e^{-iH_{-}t}|x_2\rangle$ are periodic
functions with frequencies $\omega_1$ and $\omega_2$, respectively. For $g_2\ll \omega_0$, the difference $\omega_1-\omega_2\simeq g_2$ is much smaller than each
frequency, then the 'interference' between $\langle x_3|e^{-iH_{+}t}|x_1\rangle$ and $\langle x_3|e^{-iH_{-}t}|x_2\rangle$ forms a beat with a frequency proportional to
$\omega_1-\omega_2$, so we can expect $L$ has similar behavior.

\begin{figure*}
\begin{center}
\includegraphics[width=14cm]{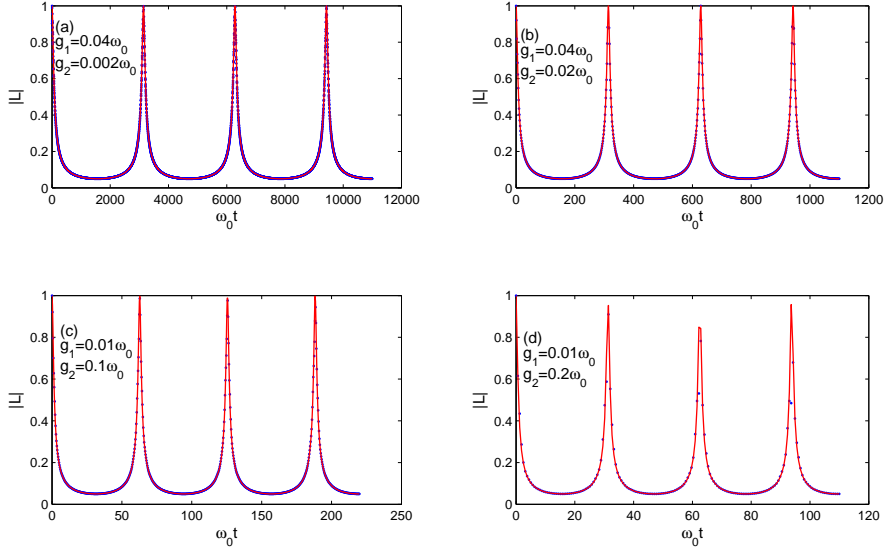}
\caption{ \label{fig1} Absolute value of the coherence $|L|$ versus the evolution time $t$, where the dots are numerical results
and the red curves are obtained from Eq.~(\ref{q2}). The figures show the comparison between numerical results and analytic formulas
with different parameters: (a), $g_1=0.04\omega_0$, $g_2=0.002\omega_0$ (b), $g_1=0.04\omega_0$, $g_2=0.02\omega_0$ (c), $g_1=0.01\omega_0$, $g_2=0.1\omega_0$ and
(d), $g_1=0.01\omega_0$, $g_2=0.2\omega_0$. The temperature is set to be $T=10\omega_0$ in all the cases.}
\end{center}
\end{figure*}

By completing the Gaussian integral, the coherence is:
\begin{eqnarray} \label{fid}
L = \frac{1}{Z\omega_0}e^{\frac{1}{2}J_i(M^{-1})_{ij}J_j}\sqrt{\frac{\omega_1\omega_2}{ |M| \sinh\beta \omega_0 \sin \omega_1t \sin \omega_2t}}
\end{eqnarray}
where the matrix:

\begin{widetext}
\begin{equation} \label{matrix}
M=\left(
        \begin{array}{ccc}
          \coth\beta \omega_0-\frac{i\omega_1}{\omega_0}\cot\omega_1t & -\frac{1}{\sinh\beta\omega_0} & \frac{i\omega_1}{\omega_0\sin\omega_1t} \\
          -\frac{1}{\sinh\beta\omega_0} & \coth\beta \omega_0+\frac{i\omega_2}{\omega_0}\cot\omega_2t & -\frac{i\omega_2}{\omega_0\sin\omega_2t} \\
          \frac{i\omega_1}{\omega_0\sin\omega_1t} & -\frac{i\omega_2}{\omega_0\sin\omega_2t} & \frac{i\omega_2}{\omega_0}\cot\omega_2t-\frac{i\omega_1}{\omega_0}\cot\omega_1t \\
        \end{array}
      \right)
\end{equation}
\end{widetext}
and $|M|=det(M)$, $J_3=J_1+J_2$. Equation (\ref{fid})
is an exact formula since no approximation has been made until now. Under the condition $g_2\ll\omega_0$,
we have $i\frac{\omega_1}{\omega_0}\cot\omega_1t\simeq i\cot\omega_1t$. By making similar approximations to all the terms containing
$\omega_1$ and $\omega_2$, we get:

\begin{equation} \label{det}
|M|\simeq \frac{2\cosh\beta\omega_0\cos(\omega_1-\omega_2)t+2i\sinh\beta\omega_0\sin(\omega_1-\omega_2)t-2}{\sinh\beta\omega_0\sin \omega_1t \sin \omega_2t}
\end{equation}

The expression of $e^{\frac{1}{2}J_i(M^{-1})_{ij}J_j}$ is still lengthy and we don't write it explicitly, but what we want to stress are (1):
$J_i(M^{-1})_{ij}J_j$ is free of divergence though $J_i$ diverges at $\omega_1t=(2n+1)\pi$ or $\omega_2t=(2n+1)\pi$ and (2): $|J_i(M^{-1})_{ij}J_j|\ll 1$, since $J_i\propto \frac{g_1}{\omega_0}$. According to these reasons $e^{\frac{1}{2}J_i(M^{-1})_{ij}J_j}\simeq 1$ and:
 \begin{equation} \label{q2}
    L\simeq \frac{1}{Z} \sqrt{\frac{1}{|M|\sinh\beta\omega_0\sin \omega_1t \sin \omega_2t}}=\frac{1}{\cos\frac{\omega_1-\omega_2}{2}t
    +i\coth\frac{\beta\omega_0}{2}\sin\frac{\omega_1-\omega_2}{2}t}
\end{equation}
Figure 1 shows the evolutions of $|L|$ with different values of parameters, where
the analytical formula Eq.~(\ref{q2}) fits with the numerical results very well
up to $g_2=0.2\omega_0$. The coherence revives at times $(\omega_1-\omega_2)T_n=2n\pi$, and near
the peaks it behaves as: $|L(t)|\simeq 1- \frac{(\omega_1-\omega_2)^2(t-T_n)^2}{2\sinh ^2(\beta\omega_0/2)}$,
so the width of every peak is $\Delta t=\frac{\sinh(\beta\omega_0/2)}{\omega_1-\omega_2}$, which decreases as the
temperature increases! For $\beta\omega_0\ll 1$ the relative error of the position of the first peak
is $\frac{\Delta t}{T_1}\simeq \frac{\beta\omega_0}{4\pi}$, so our result provides a way to measure the quadratic coupling strength precisely without
the need of cooling.

\section{Comparison with a classical oscillator}
The revival of qubit coherence is a purely quantum effect. In the quantum description, the evolution of the oscillator wavefunction separates into two branches
associated with the $|0\rangle$ and $|1\rangle$ state of the qubit, respectively: $|\varphi\rangle\rightarrow e^{-iH_\pm t}|\varphi\rangle$.
The interference of the two branches forms a beat and causes the revival of coherence. On the other hand, in classical mechanics
there is no such separation, and the coherence never revives. In the following we give a quantitatively comparison between classical and quantum theory.

For a classical oscillator, the thermal state is described by the Boltzman distribution:
$f(x,p)=\frac{\beta \omega_0}{2\pi}\exp\{-\frac {\beta \omega_0}{2}(x^2+p^2)\}$. The orbit of this oscillator with initial position $x_0$ and momentum $p_0$
is $x(t)=x_0\cos\omega_0 t+p_0\sin\omega_0 t$, and the expected value of any phase space function
$O(x(t),p(t))$ is evaluated by averaging over $x_0$, $p_0$ with respect to $f(x_0,p_0)$. For example, the correlation function
 of $x$ is:
 \begin{equation}\label{green}
    \langle x(t_2)x(t_1)\rangle= \iint x(t_2)x(t_1)f(x_0,p_0) dx_0dp_0
    =\frac{1}{\beta\omega_0}\cos\omega_0(t_2-t_1)
 \end{equation}
 which is  the Gaussian noise with spectral density $S(\omega)=\frac{1}{2\beta\omega_0}(\delta(\omega+\omega_0)+\delta(\omega-\omega_0))$.
 Now we consider the qubit coherence in two different cases:

(\textbf{1}) $g_1\neq 0$, $g_2=0$, the coherence is :
\begin{eqnarray} \label{c1}
  L_c &=& \langle e^{-2ig_1\int x(t) dt}\rangle= e^{-2g_1^2 \iint \langle x(t_2)x(t_1)\rangle dt_1dt_2} \nonumber \\
  &=&e^{-4\frac{g_1^2(1-\cos\omega_0t)}{\beta \omega_0^3}}
\end{eqnarray}
Also we know the coherence in the quantum theory is $ L=e^{-2\frac{g_1^2}{\omega_0^2}\coth\frac{\beta \omega_0}{2}(1-\cos\omega_0t)}$.
The quantum formula has similar time dependence with its classical counterpart and reduces to the latter in the high temperature limit
$\beta\rightarrow 0$, since $\coth\frac{\beta \omega_0}{2}\rightarrow \frac{2}{\beta \omega_0}$.

(\textbf{2}) $g_1=0$, $g_2\neq 0$. From the expression of $x(t)$, the coherence is evaluated as:
\begin{eqnarray} \label{c2}
  L_c = \langle e^{-ig_2\int x^2(t) dt}\rangle  
    = \frac{\beta \omega_0}{\sqrt{(\beta \omega_0+ig_2t)^2+\frac{g_2^2}{2\omega_0^2}(1-\cos 2\omega_0t)}}
   \simeq  \frac{\beta \omega_0}{\beta \omega_0+ig_2t}
\end{eqnarray}
 which decays with time and never revives.
 Clearly, the quantum mechanical result in Eq.~(\ref{q2}) does not agree with the above formula in the limit $\beta\rightarrow 0$.

So in the case of quadratic coupling, the quantum effect is largely enhanced. We know the differences between
quantum and classical theory resides in two facets: the statistics and the dynamics.
In high temperature, the quantum statistics reduces to the classical statistics, but this has nothing to do with the dynamics. Then why for linear coupling
there seems to be no difference between quantum dynamics and classical dynamics? To understand it, one notes that the coherence in the quantum theory can be written as~\cite{breuer}:
\begin{equation} \label{q}
L=\frac{1}{Z}Tr[e^{-\frac{\beta\omega_0}{2}(\hat{x}_0^2+\hat{p}_0^2)}\mathcal{T}_\rightarrow e^{-ig_1\int \hat{x}(t) dt}\mathcal{T}_\leftarrow e^{-ig_1\int\hat{x}(t) dt}]
\end{equation}
where $\hat{x}(t)=\hat{x}_0\cos\omega_0 t+\hat{p}_0\sin\omega_0 t$ is the coordinate operator in the interaction picture, $\mathcal{T}_\leftarrow$ and $\mathcal{T}_\rightarrow$
are chronological and anti-chronological time-ordering operators respectively. The point is that, the commutator
$[\hat{x}(t_1), \hat{x}(t_2)]=i\sin[\omega_0(t_2-t_1)]$ is just a $\emph{c}$ number, then one can remove the time-ordering operators
$\mathcal{T}_\leftarrow$ and $\mathcal{T}_\rightarrow$ (there will be two extra phase factors, but they cancel each other) and combine
the two time evolution operators into one: $e^{-2ig_1\int \hat{x}(t) dt}$, the expectation value of which coincides with Eq.~(\ref{c1})
in high temperature. However this is not true for the quadratic coupling model, since $[\hat{x}^2(t_1), \hat{x}^2(t_2)]$ is obviously not a $\emph{c}$ number.
Consequently, in this case the quantum dynamics is quite different from the classical dynamics.

The difference between quantum and classical dynamics also has another outcome: by recovering the Planck constant $\hbar$, we have
 $L\simeq \frac{1}{\cos\frac{g_2}{2}t+i\coth\frac{\beta\hbar\omega_0}{2}\sin\frac{g_2}{2}t}$ and $L_c\simeq  \frac{\beta \hbar\omega_0}{\beta \hbar\omega_0+ig_2t}$, it is found the two results
do not coincide in  $\hbar\rightarrow 0$. To see how this occurs, let's focus on the time ordered evolution operator, which by definition is:
\begin{equation}\label{evolution}
    \mathcal{T}_\leftarrow e^{-ig_2\int \hat{x}^2 dt/\hbar}=\lim_{\Delta t\rightarrow0} e^{-ig_2\hat{x}^2(t)\Delta t/\hbar}...e^{-ig_2\hat{x}^2(t_n)\Delta t/\hbar}e^{-ig_2\hat{x}^2(t_{n-1})\Delta t/\hbar}...e^{-ig_2\hat{x_0}^2\Delta t/\hbar}
\end{equation}
If we want to combine all these infinitesimal evolution operators into a single exponential, the Hausdorff formula tells:
\begin{equation}\label{evolution1}
     \mathcal{T}_\leftarrow e^{-ig_2\int \hat{x}^2 dt/\hbar}=e^{-ig_2\int \hat{x}^2 dt/\hbar-\frac{1}{2}g_2^2\iint[\hat{x}^2(t_1),\hat{x}^2(t_2)]dt_1dt_2/\hbar^2+...}
\end{equation}
The term $-\frac{1}{2}g_2^2\iint[\hat{x}^2(t_1),\hat{x}^2(t_2)]dt_1dt_2/\hbar^2$ can't be neglected since $[\hat{x}^2(t_1),\hat{x}^2(t_2)]\sim\hbar$
(arising from $[\hat{x},\hat{p}]=i\hbar$)
and it is of the order $\hbar^{-1}$ as the first term $-ig_2\int \hat{x}^2 dt/\hbar$, which means in this very case,
$\hat{x}$ and $\hat{p}$ can't be naively taken as $\emph{c}$ numbers even in the limit $\hbar\rightarrow0$.

\section{Influence of the environment}
In reality, the qubit-oscillator system is always under the influence of a surrounding environment. Here we model the environment with a bath of oscillators,
and the system-bath Hamiltonian is~\cite{caldeira}: $H=H_s+H_B+V$, with $H_B=\frac{1}{2}\sum_k\omega_k(x_k^2+p_k^2)$
and:
 \begin{equation}\label{2}
    V=-\sum_k c_kx_kx+\sum_k\frac{c_k^2}{2\omega_k}x^2
 \end{equation}
The spectral density of the bath is assumed to be Ohmic:
 $J(\omega)=\sum_k\frac{c_k^2}{2}\delta(\omega-\omega_k)=\frac{2\gamma}{\pi\omega_0}\omega$, where $\gamma$ is a constant,
 and the noise kernel is:
 \begin{equation}\label{noise}
    \nu(\tau)=2\int_0^{\infty}J(\omega)\coth \frac{\omega}{2kT}\cos\omega\tau=\frac{8\gamma kT}{\omega_0}\delta(\tau)\equiv\lambda\delta(\tau)
 \end{equation}

Now the coherence is: $L(t) = Tr[\rho(t)]$,
where $\rho(t)=Tr_B[e^{-i(H_{+}+H_B+V)t}\rho_0\otimes\rho_Be^{i(H_{-}+H_B+V)t}]$. After tracing over the bath,
the propagation of $\rho(t)$ has the form~\cite{caldeira}:
\begin{equation}\label{3}
    \rho(x_f,x'_f,t)=\iint J(x_f, x'_f, t; x_i, x'_i, 0)\rho(x_i,x'_i,t)dx_idx'_i
 \end{equation}
 with the propagating function:
 \begin{eqnarray}
    J(x_f, x'_f, t; x_i, x'_i, 0) &=& \int \mathcal{D}x\mathcal{D}x' e^{i\int\big[\frac{1}{2\omega_0}(\dot{x}^2-\dot{x'}^2)-\frac{1}{2}\omega_0 (x^2-x'^2)
    ++\frac{g_2}{2}(x^2+x'^2)-\frac{\gamma}{\omega_0}(\dot{x}+\dot{x'})q+\frac{i\lambda}{4}(x-x')^2]dt} \nonumber \\
    &=&\int \mathcal{D}r\mathcal{D}q e^{i\int\big[\frac{1}{2\omega_0}\dot{r}\dot{q}-\frac{1}{2}\omega_0 rq
    -\frac{g_2}{4}(r^2+q^2)-\frac{\gamma}{\omega_0}\dot{r}q+\frac{i\lambda}{4}q^2]dt} \nonumber \\
    &\equiv & \int \mathcal{D}r\mathcal{D}q e^{iA\{r(\tau),q(\tau)\}}
 \end{eqnarray}
and $r=x+x'$, $q=x-x'$, $A\{r(\tau),q(\tau)\}\equiv\int\big[\frac{1}{2\omega_0}\dot{r}\dot{q}-\frac{1}{2}\omega_0 rq
-\frac{g_2}{4}(r^2+q^2)-\frac{\gamma}{\omega_0}\dot{r}q+\frac{i\lambda}{4}q^2\big]d\tau$. The path integral
is evaluated by the stationary phase method, giving: $J(x_f, x'_f, t; x_i, x'_i, 0)=N(t)e^{iA_m (r_f, q_f, r_i ,q_i, t)}$,
where $A_m$ is the extremum of $A$ that depends on the boundary conditions and $N(t)$ is the Gaussian integral of the fluctuations.
The extremum path is determined by $\frac{\delta A}{\delta r}=\frac{\delta A}{\delta q}=0$:
\begin{eqnarray} \label{eigen}
  \ddot{r}+2\gamma\dot{r}+\omega^2_0r+(g_2-i\lambda)q &=& 0  \nonumber \\
  \ddot{q}-2\gamma\dot{q}+\omega^2_0q+g_2r &=& -2g_1\omega_0
\end{eqnarray}
Generally, the solution to these equations is a linear combination of $e^{z_i\tau}$ ($i=1,2,3,4$) and a constant (due to the term
$-2g_1\omega_0$), where:
\begin{eqnarray}
  z_{1,2} &=& \pm \sqrt {2\gamma^2-\omega^2_0+\sqrt{4\gamma^4-4\gamma^2\omega^2_0+g^2_2\omega^2_0-ig_2\lambda\omega^2_0}} \nonumber \\
  z_{3,4} &=& \pm \sqrt {2\gamma^2-\omega^2_0-\sqrt{4\gamma^4-4\gamma^2\omega^2_0+g^2_2\omega^2_0-ig_2\lambda\omega^2_0}}
\end{eqnarray}
By substituting the solution to $A$, one easily finds that $A_m$ is a linear combination of $e^{(z_i+z_j)t}$.
If there is no bath, we have $\gamma=\lambda=0$ and $z_{1,2}=\pm i\sqrt{\omega_0(\omega_0-g_2)}$, $z_{3,4}=\pm i\sqrt{\omega_0(\omega_0+g_2)}$,
then every term of $e^{(z_i+z_j)t}$ is a periodic function of $t$, and so is $A_m$ with a frequency $\propto|z_3-z_1|\simeq g_2$. In the presence of the bath,
 however, $z_i$ has both real and imaginary parts and $A_m$ is not a periodic function, so one can only expect that,
 under certain conditions, the revival of coherence can be reached approximately in a time scale $2\pi/g_2$.
This requires the amplitude of $e^{(z_i+z_j)t}$ varies slightly in such a period, which means:
$Re(z_i)/g_2\ll1$ ($Re$ indicates the real part). According to (\ref{eigen}), it is equivalent to: $\gamma\ll g_2$,  $\lambda \ll g_2$,
which indicates the influence of the bath is negligible in a duration $\sim2\pi/g_2$.
If these conditions can't be met, the coherence never revives, and its evolution will be similar to the classical result
presented in Sec. II.

\section{Summary and Discussion}
In summary, we studied the dynamics of a qubit-oscillator model and gave the analytical formula for the qubit coherence. In principle,
this formula provides a way to measure the quadratic coupling strength $g_2$.
An interesting question is, for a system with both linear coupling and quadratic coupling,
to what extent the linear approximation is reasonable. From Eqs. (\ref{fid}) and (\ref{matrix}), it's seen that for $g_2t\ll1$ we have
: $\omega_1t\simeq\omega_2t\simeq\omega_0t$, then $\frac{1}{2}J_i(M^{-1})_{ij}J_j\simeq -2\frac{g_1^2}{\omega_0^2}\coth\frac{\beta \omega_0}{2}(1-\cos\omega_0t)$
and Eq.(\ref{fid}) has a simple form:
\begin{eqnarray} \label{coherence}
L\simeq\frac{1}{1+ig_2t\coth\frac{\beta\omega_0}{2}}e^{-2\frac{g_1^2}{\omega_0^2}\coth\frac{\beta \omega_0}{2}(1-\cos\omega_0t)} \nonumber \\
\end{eqnarray}
where the exponential is just the result of linear coupling model. So the validity of the linear model demands $g_2t\coth\frac{\beta\omega_0}{2}\ll 1$,
which means short time and low temperature, otherwise the quadratic coupling must be taken into account.

\begin{figure}
\begin{center}
\includegraphics[width=10cm]{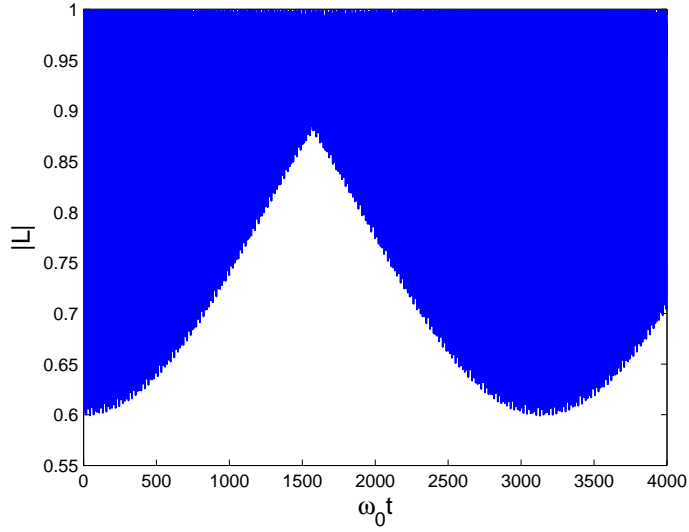}
\caption{ \label{fig2} The coherence $|L|$ versus the evolution time $t$ under Hahn-echo. The parameters are:
$g_1=0.04\omega_0$, $g_2=0.004\omega_0$ and $T=10\omega_0$. The shadow is the fast oscillation of $|L|$
with frequency $\omega_0/2$, and $g_2$ modulates this oscillation.}
\end{center}
\end{figure}

In experiments the Hahn-echo~\cite{hahn} is often applied to suppress the thermal noise of the environment. This
is performed by flipping the spin at a moment $t/2$: $C_0|0\rangle+C_1|1\rangle\rightarrow C_0|1\rangle+C_1|0\rangle$,
then measuring the coherence at time $t$. The coherence in this case is:
\begin{equation}
L =\frac{1}{Z}\int dx_1 dx_2  \langle x_1|e^{-\beta H_0}|x_2\rangle\langle x_2|e^{iH_{-}t/2}e^{iH_{+}t/2}e^{-iH_{-}t/2}e^{-iH_{+}t/2}|x_1\rangle
\end{equation}
By using the same method in Section I, it is evaluated as:
\begin{eqnarray}
|L|\simeq \exp \{-4\frac{g_1^2}{\omega_0^2}\coth\frac{\beta\omega_0}{2} (\cos\frac{\omega_0t}{2}-\cos\frac{g_2t}{4})^2 \}
\end{eqnarray}

Figure 2 shows the evolution of $|L|$ under Hahn-echo. The oscillation of $|L|$ is modulated by the quadratic coupling,
one can show the lower bound of $|L|$ is $e^ {-4\frac{g_1^2}{\omega_0^2}\coth\frac{\beta\omega_0}{2} (1+|\cos\frac{g_2t}{2}|)^2 }$.
The coherence of linear coupling model under Hahn echo is obtained by simply setting $g_2=0$
in the above equation, so the effect of quadratic coupling is not so remarkable as in the case
of free evolution, as long as $g_2t\ll1$ the validity of linear approximation can be guaranteed.

An important generalization to our model is that the qubit Hamiltonian has the form: $H_q=\frac{1}{2}\omega_q\sigma_z+\frac{1}{2}\Delta\sigma_x$,
which is common for a superconducting qubit~\cite{tsai,shnirman1,ithier,luca,wei,reuther}. In this case the coupling to an oscillator
will lead to dissipation as well dephasing of the qubit. For $\Delta\ll g_2\ll\omega_q$, the rotation wave approximation can be applied
to eliminate the $\sigma_x$ term~\cite{shnirman3} and the previous result is valid in a time scale $2\pi/g_2$.
For larger $\Delta$ the situation is much more complicated, and a possible way is to use the perturbation
approach as in ~\cite{luca}. Nevertheless, it remains an open problem. Furthermore, if the qubit-oscillator system is  weakly coupled to a bath with
short correlation time, the Bloch-Redfield approach can be applied to get the decoherence rate of the qubit, as is done in~\cite{ithier}.

\acknowledgements
We are very grateful to {\L}. Cywi{\'n}ski for his helpful discussion.
This work is supported by NKBRP (973 Program) 2014CB848700 and 2016YFA0301200, NSFC No. 11374032 and NSAF U1530401.


\begin{thebibliography}{99}
\bibitem{zurek1} W. H. Zurek, Rev. Mod. Phys. {\bf 75}, 715 (2003).
\bibitem{everett} H. Everett III, Rev. Mod. Phys. {\bf 29}, 454 (1957).
\bibitem{lindblad} G. Lindblad,  Commun. Math. Phys. {\bf 48}, 119 (1976).
\bibitem{leggett1} A. O. Caldeira and A. J. Leggett, Phys. Rev. A. {\bf 31}, 1059 (1985).
\bibitem{leggett2} A. J. Leggett, S. Chakravarty, A. T. Dorsey, Matthew P. A. Fisher, Anupam Garg, and W. Zwerger,
Rev. Mod. Phys. {\bf 59}, 1 (1987).
\bibitem{unruh} W. G. Unruh and W. H. Zurek,  Phys. Rev. D. {\bf 40}, 1071 (1989).
\bibitem{myatt} C. J. Myatt, B. E. King, Q. A. Turchette, C. A. Sackett, D. Kielpinski, W. M. Itano,
                 C. Monroe,  and D. J. Wineland,  Nature. {\bf 403}, 269 (2000).  -
\bibitem{zurek2} W. H. Zurek, Phys. Today. {\bf 44}, 36 (1991).
\bibitem{weiss} U. Weiss, {\em Quantum Dissipative Systems} (World Scientific, Singapore, 1999).
\bibitem{zanardi} P. Zanardi and M. Rasetti, Phys. Rev. Lett. {\bf 79}, 3306 (1997).
\bibitem{shor} P. W. Shor, Phys. Rev. A. {\bf 52}, R2493 (1995).
\bibitem{ekert} A. Ekert and C. Macchiavello, Phys. Rev. Lett. {\bf 77} 2585 (1996).
\bibitem{lloyd1} L. Viola and S. Lloyd, Phys. Rev. A {\bf 58}, 2733 (1998).
\bibitem{lloyd2} L. Viola, E. Knill and S. Lloyd, Phys. Rev. Lett. {\bf 82}, 2417 (1999).
\bibitem{zhao1} N. Zhao, J. Honert, B. Schmid, M. Klas, J. Isoya, M. Markham,
D. Twitchen, F. Jelezko, R.-B. Liu, H. Fedder, and J.Wrachtrup, Nat. Nanotechnol. {\bf 7}, 657 (2012).
\bibitem{lukin1} S. Kolkowitz, Q. P. Unterreithmeier, S. D. Bennett, and M. D.
Lukin, Phys. Rev. Lett. {\bf 109}, 137601 (2012).
\bibitem{hanson} T. H. Taminiau, J. J. T. Wagenaar, T. van der Sar, F. Jelezko,
    V. V. Dobrovitski, and R. Hanson, Phys. Rev. Lett. {\bf 109}, 137602 (2012).
\bibitem{lavrik} N. V. Lavrik and P. G. Datskos, Appl. Phys. Lett. {\bf 82}, 2697 (2003).
\bibitem{ganzhorn} M. Ganzhorn, S. Klyatskaya, M. Ruben, and W. Wernsdorfer, Nat. Nanotechnol. {\bf 8}, 165 (2013).
\bibitem{chaste} J. Chaste, A. Eichler, J. Moser, G. Ceballos, R. Rurali,
and A. Bachtold, Nat. Nanotechnol. {\bf 7}, 301 (2012).
\bibitem{jensen} K. Jensen, K. Kim, and A. Zettl, Nat. Nanotechnol. {\bf 3}, 533 (2008).
\bibitem{naik} A. K. Naik, M. S. Hanay, W. K. Hiebert, X. L. Feng, and M. L. Roukes, Nat. Nanotechnol. {\bf 4}, 445 (2009).
\bibitem{zhao2} N. Zhao and Z. Q. Yin, Phys. Rev. A. {\bf 90}, 042118 (2014).
\bibitem{uhrig}  G. S. Uhrig, Phys. Rev. Lett. {\bf 98}, 100504 (2007).
\bibitem{lidar} K. Khodjasteh and D. A. Lidar, Phys. Rev. A. {\bf 75}, 062310 (2007).
\bibitem{sarma} {\'L}. Cywi{\'n}ski, R. M. Lutchyn, C. P. Nave, and S. Das Sarma, Phys. Rev. B. {\bf 77}, 174509 (2008).
\bibitem{uys} M. J. Biercuk, A. C. Doherty, and H. Uys, J. Phys. B: At. Mol. Opt. Phys. {\bf 44}, 154002 (2011).
\bibitem{liu} W. Yang, Z.-Y. Wang, and R.-B. Liu, Front. Phys. {\bf 6}, 2 (2011).
\bibitem{viola} K. Khodjasteh, J. Sastrawan, D. Hayes, T. J. Green, M. J. Biercuk, and L. Viola, Nat. Commun. {\bf 4}, 2045 (2013).
\bibitem{itano} M. J. Biercuk, H. Uys, A. P. VanDevender, N. Shiga, W. M. Itano, and J. J. Bollinger, Nature (London) {\bf 458}, 996 (2009).
\bibitem{tsai} F. Yoshihara, K. Harrabi, A. O. Niskanen, Y. Nakamura, and
J. S. Tsai, Phys. Rev. Lett. {\bf 97}, 167001 (2006).
\bibitem{shnirman1} K. Kakuyanagi, T. Meno, S. Saito, H. Nakano, K. Semba,
H. Takayanagi, F. Deppe, and A. Shnirman, Phys. Rev. Lett. {\bf 98}, 047004 (2007).
\bibitem{shnirman2} A. Shnirman, Y. Makhlin, and G. Sch\"{o}n, Phys. Scr. T {\bf 102}, 147 (2002).
\bibitem{vion} D. Vion, A. Aassime, A. Cottet, P. Joyez, H. Pothier, C. Urbina,
D. Esteve, and M. H. Devoret, Science {\bf 296}, 886 (2002).
\bibitem{medford} J. Medford, J. Beil, J. M. Taylor, E. I. Rashba, H. Lu, A. C.
Gossard, and C. M. Marcus, Phys. Rev. Lett. {\bf 111}, 050501 (2013).
\bibitem{ithier} G. Ithier, E. Collin, P. Joyez, P. J. Meeson, D. Vion, D. Esteve,
F.Chiarello, A. Shnirman, Y.Makhlin, J. Schriefl, and G. Sch\"{o}n, Phys. Rev. B {\bf 72}, 134519 (2005).
\bibitem{shnirman3} Y. Makhlin and A. Shnirman, Phys. Rev. Lett. {\bf 92}, 178301 (2004).
\bibitem{lukasz} {\L}. Cywi{\'n}ski, Phys. Rev. A {\bf 90}, 042307 (2014).
\bibitem{luca} L. Chirolli and G. Burkard, Phys. Rev. B {\bf 80}, 184509 (2009).
\bibitem{wei}  Jiang Wei, Yu Yang and Wei Lian-Fu, Chin. Phys. B {\bf 20}, 080307 (2011).
\bibitem{reuther} G. M. Reuther, D. Zueco, P. Hanggi and S. Kohler, New J. Phys. {\bf 13}, 093022 (2011).
\bibitem{yang} W. Yang and R.-B. Liu, Phys. Rev. B {\bf 78}, 085315 (2008).
\bibitem{breuer} H. P. Breuer, F. Petruccione, {\em The Theory of Open Quantum Systems} (Oxford University Press, 2002).
\bibitem{caldeira} A. O. Caldeira and A. J. Leggett, Physica A. {\bf 121}, 587 (1983).
\bibitem{hahn} E. L. Hahn, Phys. Rev. {\bf 80}, 580 (1985).
\bibitem{lloyd} L. Viola, E. Knill, and S. Lloyd, Phys. Rev. Lett. {\bf 82}, 2417 (1999).
\bibitem{unruh1} W. G. Unruh, Phys. Rev. A {\bf 51}, 992 (1995).
\end{thebibliography}
\end{document}